\documentclass[onecolumn,12pt,draftcls]{IEEEtran}

\usepackage{times}
\usepackage{graphicx}
\usepackage{fancyhdr}   
\usepackage{amsmath}
\usepackage{amssymb}
\usepackage{dsfont}
\usepackage{flushend}
\usepackage{bm}
\usepackage{url}
\usepackage{amsfonts}
\usepackage{dsfont}
\usepackage{amscd}
\usepackage{psfrag}
\usepackage{cases}
\usepackage{subfigure}

\usepackage{amsfonts}
\usepackage{graphicx, subfigure}
\usepackage{xcolor}
\usepackage{stfloats}
\usepackage{algorithm}
\usepackage{algpseudocode}

\newtheorem{defn}{Definition}[section]

\DeclareMathOperator*{\argmin}{arg\,min}
\DeclareMathOperator*{\argmax}{arg\,max}

\begin{document}

\title{Throughput and Delay Analysis in Video Streaming over Block-Fading Channels}
\author{\IEEEauthorblockN{G. Cocco$^{\ddag}$, D. G\"{u}nd\"{u}z$^{*}$ and C. Ibars$^{\dag}$  }\\
\IEEEauthorblockA{$^{\ddag}$\small German Aerospace Center (DLR), We{\ss}ling, Germany}\\
\IEEEauthorblockA{$^{*}$\small Imperial College, London, United Kingdom}\\
\IEEEauthorblockA{$^{\dag}$\small Intel Corporation, Santa Clara, CA, USA}\\
\small{giuseppe.cocco@dlr.de, d.gunduz@imperial.ac.uk, christian.ibars.casas@intel.com}
}
\date{}
\maketitle

\begin{abstract}
We study video streaming over a slow fading wireless channel. In a streaming application video packets are required to be decoded and displayed in the order they are transmitted as the transmission goes on. This results in per-packet delay constraints, and the resulting channel can be modeled as a physically degraded fading broadcast channel with as many virtual users as the number of packets. In this paper we study two important quality of user experience (QoE) metrics, namely \textit{throughput} and \textit{inter-decoding delay}. We introduce several transmission schemes, and compare their throughput and maximum inter-decoding delay performances. We also introduce a genie-aided scheme, which provides theoretical bounds on the achievable performance. We observe that adapting the transmission rate at the packet level, i.e., periodically dropping a subset of the packets, leads to a good tradeoff between the throughput and the maximum inter-decoding delay. We also show that an approach based on initial buffering leads to an asymptotically vanishing packet loss rate at the expense of a relatively large initial delay. For this scheme we derive a condition on the buffering time that leads to throughput maximization.
\end{abstract}
\vspace{-.1in}
\section{Introduction}\label{sec:intro}
Video traffic constitutes a large portion of today's Internet data flow, and it is foreseen to exceed $70\%$ of the total IP traffic within the next five years \cite{cisco_forecast_2014}. A significant portion of the video traffic is generated  by streaming applications, such as YouTube and Netflix. This, together with the increasing utilization of mobile terminals for streaming high-definition video content, poses growing challenges to mobile network operators in terms of bandwidth availability and quality of user experience (QoE).


Mobile wireless channels are often modelled with block fading, where the channel gain stays constant during the channel coherence time, and changes independently across channel blocks according to a certain probability distribution \cite{ozarow_TVT_mobile_radio}. From the extensive literature on fading channels (see, e.g., \cite{goldsmith_TIT1997_capacity_fading}-\nocite{caire_TIT91_power_control}\nocite{berry_gallagher_TIT_2002}\nocite{zhang_jsac_2010_video_fading}\nocite{Hanly:IT:98}\nocite{Tse:book}\cite{gong_twc_2014_delay_video}), it emerges that a pivotal role for reliable communications is played by the delay constraint, which is a critical design parameter in streaming applications.

In \cite{ng_gunduz_2007_recursive_power} and \cite{gunduz_2008_joint_source} the broadcast strategy proposed in \cite{Shamai:ISIT:97} is used to improve the end-to-end quality in multimedia transmission. However, the broadcast strategy requires encoding bits into multiple superposed messages of increasing rates, and this level of fine adaptation is not possible in practical multimedia communication systems, in which the encoding rate is fixed by a higher layer application\footnote{Some streaming protocols, such as HTTP Live Streaming, allow rate adaption among only a limited number of available rates.}\cite{drapernework2006benefitsstreaming_fading}. Moreover, practical network architectures are strictly layered, and the channel encoder is typically oblivious to the video coding scheme used by the application layer; and therefore, rate adaptation is usually not possible at the code level. Video packets received by the channel encoder are already video-encoded at a fixed rate, which cannot be changed. On the other hand, the channel encoder can choose to drop some of the video packets, and achieve rate adaptation at the packet level at the expense of \textit{inter-decoding delay} at the receiver.

In the Moving Picture Experts Group (MPEG) standard, the video encoder output units are called group of pictures (GOP). Each GOP consists of an I- frame and a number of P- and B-frames \cite{wang_video_proc_and_comm}. A GOP can be decoded and displayed independently of the previous and following GOPs. We assume that a whole GOP (or an integer number of GOPs) forms one video packet, and the coding rate is normalized such that the display time of a GOP (or an integer number of GOPs) is equal to the channel coherence time\footnote{With this we implicitly assume a slow varying channel, for example, a mobile terminal moving at pedestrian speed.}.

We consider streaming over a Gaussian block fading channel, in which the transmitter has no channel state information (CSIT), which is the case for networks with large round trip delay (like satellite networks), or wireless broadcast networks with a large number of users\footnote{In the downlink channel with many receiving terminals, acquisition of CSIT is not viable, since this requires the transmission of an extensive amount of information which may result in the \emph{feedback implosion} problem \cite{sali2007_feedback_implosion}.}. Due to the lack of CSIT, the transmitter uses a fixed transmission rate. In order to minimize the probability of packet loss over the channel, the transmission rate is kept at the minimum value that allows no freezing in the display process at the receiver provided no packet is lost. This implies that the transmission time of a packet is equal to its display time (assuming that the time needed to process the packet at the receiver is negligible), which is assumed to be constant for all packets. In the streaming scenario, this imposes a different decoding deadline for each video packet, i.e., the first packet needs to be received by the end of the first channel block, the second packet by the end of the second block, and so on. Modeling the decoder at each channel block as a distinct virtual receiver, this channel can be seen as a physically degraded fading broadcast channel with as many virtual users as the number of channel blocks.

The loss of a data packet implies the loss of the corresponding GOP; and hence, an interruption in the playback of the video at the end user, which lasts until the next packet is received. In \cite{video_quality_assessment_JSTSP_2012} the quality degradation due to GOP losses as perceived by the end user has been assessed by streaming pre-recorded videos while introducing video segment losses in a controlled fashion. The results illustrate that users are more tolerant to long freezes with respect to choppy playback, that is, few long freezing events are on average preferred to many short freezing events. However, this is no longer true if the transmission is for a live event, such as a sport event or news video. In this case, the loss of a large chunk of video content, which may lead to loss of important information, is much worse than choppy playback quality. In this paper we target the latter kind of video content, and consider the interdecoding delay as a performance measure.

The effect of GOP loss in video streaming has been studied in \cite{lin_tip2010_model_video_loss}, \cite{liang_2008_effect_packet_loss} and \cite{huszak_isccsp2010_gop_loss_effect}. In the video streaming literature, the problem is usually tackled at the network level, focusing on the effect of packet loss rate, delay and jitter \cite{ITU_T_Y1541}. However, these parameters are usually assumed to be given as fixed values to the system designer, or studied from a networking perspective, where packet losses are mainly due to buffer overflow, while jitter is due to the congestion level of the network, link failures and dynamic routing.  The problem of radio resource allocation in wireless multimedia transmission over frequency selective channels is studied in \cite{toni_JSAC2012_ch_cod_optim_video_str} and \cite{zhao_glob_2010_video_str_ofdm}.

We study the interaction between the physical layer and the display process of the received video data. In particular, we study different communication strategies, each of which adopts a different policy to select the subset of messages to be transmitted, as well as the amount of resources (in terms of transmission time) dedicated to each message, which has an impact on the successful decoding probability. The performance of these strategies is evaluated based on two figures of merit: average throughput and maximum inter-decoding delay \cite{zeng_2012:joint_coding}. The interaction between the display process and the lower layers is of fundamental importance for streaming services such as Dynamic Adaptive Streaming over HTTP (DASH), that need an estimation of the link quality in order to provide an adequate QoE to the end users. In its current implementation DASH uses the information about the link status at each user in order to optimize the QoE that can be provided with the available resources \cite{Aparicio-Pardo:2015:TLA:2713168.2713177}. However, DASH systems require a feedback link that instructs the transmitter on the highgest bit-rate that can be received in the current channel condition, whereas we assume no information on the current channel state at the transmitter, and thus the optimisation of the transmission strategy at the transmitter has to be done independently of the current channel condition.


While there is an extensive literature on the higher layer analysis of video streaming applications \cite{van_der_schaar_2003}, research on the physical layer aspects of streaming focus mostly on code construction \cite{Bogino:ISCAS:07}, \cite{Badr:Globecom:10}, \cite{Leong:ISIT:12}. The diversity-multiplexing trade-off for a streaming system is studied in \cite{khisti_11_IT}. The channel model we study here is the dual of the streaming transmitter model studied in \cite{cocco11_real_time_BC}, \cite{cocco13_twc_streaming}, where the data packets, rather than being available at the transmitter in advance and having a per-packet delay constraint, arrive gradually over time, and have a global delay constraint.

We propose four different transmission schemes based on time-sharing\footnote{Part of the present work has been presented in \cite{cocco_icc_2013}.}. More elaborate transmission techniques have been previously studied in literature such as in \cite{ng_gunduz_2007_recursive_power}. In \cite{toni_TIP2009} the problem of  still images transmitting over slow fading channel using a FEC-based multiple description encoder over an OFDM modulation was studied. Unlike in such previous works, we exclusively focus on time-sharing transmission because of its applicability in practical systems, as it leads to lower complexity decoding schemes with respect to, for example, successive interference cancellation, which is required in the case of superposition transmission. Moreover, the throughput and delay analysis is not completely understood even for this relatively simpler transmission scheme. In particular, we consider \emph{memoryless transmission (MT)}, \emph{equal time-sharing (eTS)}, \emph{pre-buffering (PB)} and \emph{windowed time-sharing (wTS)} schemes. We also consider an informed transmitter (IT) bound on the achievable throughput and delay performances, assuming perfect CSIT. We compare these achievable schemes and the informed transmitter bound in terms of both throughput and maximum inter-decoding delay. Our results provide fundamental performance bounds as well as an insight for the design of practical video streaming systems over wireless fading channels.

The rest of the paper is organized as follows. In Section \ref{sec:sysmod} we present the system model. In Section \ref{sec:upper_bound} we derive informed transmitter bounds on throughput and average maximum delay. In Section \ref{sec:enco_schemes} we presents four different transmission schemes and, for each of them, we analyze throughput and delay. Section \ref{sec:num_res} contains the numerical results, while the conclusions are drawn in Section \ref{sec:conclusions}.
\vspace{-.1in}
\section{System Model}\label{sec:sysmod}
We consider a video streaming system over a block fading channel. The channel is constant for a block of $n$
channel uses and changes in an independent and identically distributed (i.i.d.) manner from one block to the next. We assume
that the file to be streamed to the receiver consists of $M$ independent packets denoted by $W_1,\ldots,W_M$, all available at the transmitter at the very beginning. The receiver wants to decode these packets gradually as the transmitter continues its transmission. We assume that the packet $W_t$ needs to be decoded by the
end of channel block $t$, $t=1, \ldots, M$, otherwise it becomes useless. The data packets all have the same size; and it is assumed that each packet is generated at
rate $R$ bits per channel use (bpcu), which is fixed by the application layer, i.e., $W_t$ is chosen randomly
with uniform distribution from the set $\mathcal{W}_t =
\{1,\ldots,2^{nR}\}$ \cite{thomas_infotheo}. The channel in block $t$ is given by
\[
\mathbf{y}[t] = h[t] \mathbf{x}[t] + \mathbf{z}[t],
\]
where $h[t]$ is the channel state, $\mathbf{x}[t]$ is the length-$n$
channel input vector, $\mathbf{z}[t]$ is a vector of i.i.d. zero mean unit-variance
Gaussian noise, and $\mathbf{y}[t]$ is the length-$n$ channel output
vector at the receiver. Instantaneous channel states are known only at
the receiver, while the transmitter has only statistical channel knowledge, i.e., it knows the probability density function (pdf) of $h(t)$. We have a short-term average power
constraint of $P$, i.e., $\mathrm{E}[\mathbf{x}[t]\mathbf{x}[t]^\dag] \leq
nP$ for $t=1, \ldots, M$, where $\mathbf{x}[t]^\dag$ represents the
Hermitian transpose of $\mathbf{x}[t]$.
\begin{figure}
\centering
\includegraphics[width=3.5in]{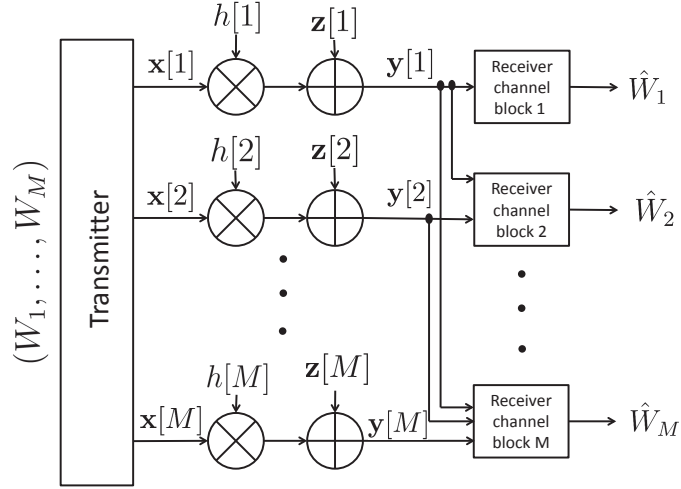} \caption{\footnotesize Equivalent channel
model for streaming a video file composed of $M$ packets over $M$
blocks of the fading channel to a single receiver with a per packet delay constraint.}
\label{fig:message}
\end{figure}

The channel from the source to the receiver can be seen as a
physically degraded broadcast channel, such that the decoder at each channel block
acts as a virtual receiver trying to decode the packet corresponding to that channel block. See Fig. \ref{fig:message} for an illustration of
this channel model. We denote the instantaneous
channel capacity over channel block $t$ by $C_t$:
\begin{eqnarray}\label{eqn:capacity}
C_t \triangleq \log_2(1+ \phi[t]P),
\end{eqnarray}
where $\phi[t]=|h[t]|^2$ is a random variable distributed according to a zero-mean pdf $f_{\Phi}(\phi)$. We define $\overline{C} \triangleq E\{C_t\}$, $E\{x\}$ being the mean value of $x$.

We define the average throughput, $\overline{T}$, as the average decoded rate at the end of $M$ channel blocks:
\begin{eqnarray}\label{eqn:throughput}
\overline{T}\triangleq\frac{R}{M}\sum_{m=1}^M m\cdot\eta(m),
\end{eqnarray}
where $\eta(m)$ is the probability of decoding exactly $m$ messages out of $M$.

In addition to the average throughput, we also study the \textit{frame delay}, which is defined as the maximum number of consecutive channel blocks
in which the corresponding message is not decoded, denoted by $D^{\text{max}}$. When a video packet over a channel block is not decoded at the receiver, video playback at the receiver's device stalls, and the user continues to see the same video frame until a new GOP is successfully received. Since $D^{\text{max}}$ is also a random variable whose realization depends on the channel, we consider the \emph{average maximum delay} $\overline{D}^{\text{max}}$ as our performance measure. We have:
\begin{align}\label{eqn:delay_def_sysmod}
\overline{D}^{\text{max}} \triangleq \sum_{d=1}^M d \cdot
Pr\{\text{$D^{\text{max}}=d$}\}=\sum_{d=1}^M
Pr\{\text{$D^{\text{max}}\geq d$}\}.
\end{align}
In the next section, we first study an informed transmitter bound on the system performance, assuming perfect CSIT about all the future channel realizations.

\vspace{-.1in}
\section{Informed Transmitter Bound}\label{sec:upper_bound}
An upper bound on the achievable average throughput and a lower bound on the average maximum inter-decoding
delay can be obtained by assuming that the transmitter is informed about the exact
channel realization over all the $M$ channel blocks non-causally. This allows the
transmitter to optimally allocate the available resources among the messages. In particular, knowing the channels
\emph{a priori} the transmitter can choose the optimal subset
$S_{\text{opt}}$ of messages to be transmitted that maximizes $\overline{T}$ and minimizes $\overline{D}^{\text{max}}$. Note that power allocation across channel blocks is not possible due to short-term power constraint. In order to find the set of messages $S_{\text{opt}}$ that minimizes the average maximum delay, we first find the maximum number of decodable messages for the given channel realizations. It follows from the physically degraded broadcast channel model depicted in Fig. \ref{fig:message} that the total number of
messages that can be decoded up to channel block $t$, denoted by $\Psi(t)$, $t=1,\ldots,M$,
is bounded as:
\begin{eqnarray}\label{eqn:lb_informed}
\Psi(t)\leq\min\left\{t,\left\lfloor \frac{I^{\text{tot}}(t)}{R}\right\rfloor\right\},\end{eqnarray}
where $I^{\text{tot}}(t)\triangleq\sum_{i=1}^{t}C_i,$ is the total mutual
information (MI) accumulated up to and including channel block $t$, while $\lfloor x\rfloor$ is the largest integer smaller than or equal to $x$.
\begin{figure}[h!]
\centering
\includegraphics[width=3in]{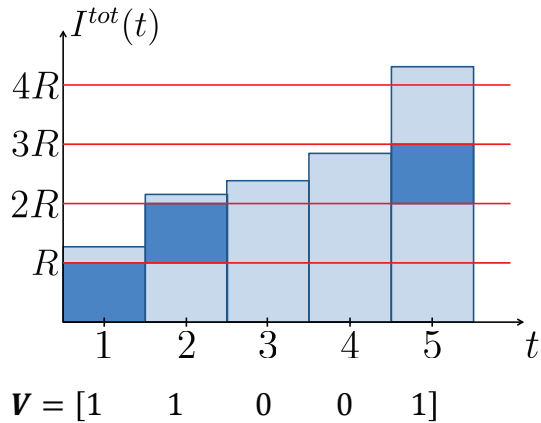}
\caption{$I^{\text{tot}}(t)$ plotted against $t$, and the corresponding vector $\mathbf{V}$ in case of throughput-optimal transmission. The light blue bars represent the amount of MI accumulated in each of the $5$ channel blocks considered, while the dark blue rectangles indicate a decoding event and represent the amount of MI that is used to decode a message.}\label{fig:allocation_1}
\end{figure}
At each channel block $t$, we check whether we can decode packet $W_t$ in addition to the packets that have already been decoded. Note that there is no gain in decoding a packet prior to its decoding deadline. Let $v(t) \in \{0, 1\}$ denote the decoding event for $W_t$, i.e., $v(t)=1$, if $W_t$ is decoded, and $v(t)=0$ if not. We have $\Psi(t)=v(1)+\cdots+v(t)$, and
\begin{align}\label{eqn:lb_algo}
v(t+1)=
\begin{cases}
1 & \text{ if } I^{\text{tot}}(t+1)\geq \left(\Psi(t)+1\right)R, \\
0 & \text{ otherwise}.
\end{cases}
\end{align}
This recursion returns the $M$-length binary vector $\mathbf{V}=[v(1) \cdots v(M)]$, which corresponds to a transmission scheme that maximizes the throughput. Although $\mathbf{V}$ represents an optimal solution in terms of $\overline{T}$, it may be suboptimal in terms of $\overline{D.}^{\text{max}}$ From the maximum delay perspective it may be a better choice not to transmit some of the packets even if enough mutual information could be accumulated by their deadlines, and instead to transmit packets that are further in the sequence. This is equivalent to shifting rightwards some of the $1$'s in $\mathbf{V}$ so that the number of consecutive $0$'s in the vector is minimized. Note that this process leaves the throughput unchanged.

Let us consider the example shown in Fig. \ref{fig:allocation_1}, where the mutual information accumulated by the receiver at the end of channel block $t$, $I^{\text{tot}}(t)$ is plotted against the channel block number. The lines $I^{\text{tot}}(t)=jR$, $j=1,\ldots, 4$, indicate the threshold values of $I^{\text{tot}}(t)$ after which a new message can be decoded. The vector $\mathbf{V}$ has entries equal to $1$ in correspondence to decoding events (shadowed areas) and zero in correspondence to channel blocks in which the receiver does not decode the corresponding message.

\begin{figure}[h!]
\centering
\includegraphics[width=3in]{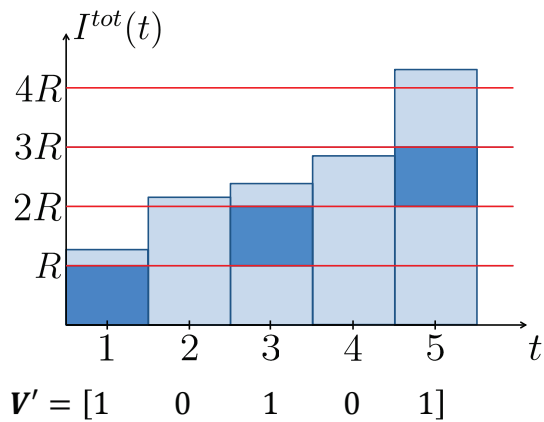}
\caption{$I^{\text{tot}}(t)$ plotted against $t$, and the corresponding vector $\mathbf{V}$ in case of throughput- and delay-optimal transmission. The light blue bars represent the amount of MI accumulated in each of the $5$ channel blocks considered, while the dark blue rectangles indicate a decoding event and represent the amount of MI that is used to decode a message.}\label{fig:allocation_2}
\end{figure}
With reference to Fig. \ref{fig:allocation_1}, the iterative process described by Eqn. (\ref{eqn:lb_algo}) returns the sequence $\mathbf{V}=[1 1 0 0 1]$. This allocation achieves a throughput of $3/5$ and a maximum delay of $2$. However, a better choice for the transmitter is to transmit message $W_3$ instead of $W_2$, as shown in Fig. \ref{fig:allocation_2}. This gives the new allocation $\mathbf{V}'=[1 0 1 0 1]$, which has the same throughput as $\mathbf{V}$ but a maximum delay of $D^{max}=1$ instead of $2$.

In order to minimize the maximum delay, the transmitter can choose to drop a message even if it could be decoded with high probability. In other words, the resources are allocated to a message with a higher index, which, if decoded, would lead to a lower maximum delay. Note that the maximum delay is optimized without decreasing the average throughput. Next we provide the necessary definitions and results to introduce the algorithm \texttt{Min\_Del\_Max\_Rate}, which optimizes both $\overline{T}$ and $\overline{D}^{\text{max}}$.\\

\begin{defn} Let $\mathbf{V}_{\text{lb},D}$ denote the binary string of length $M$ with maximum number of consecutive zeros equal to $D$, which has the smallest number of $1$'s and the smallest decimal representation.\\
\end{defn}

If $M>D$, $\mathbf{V}_{\text{lb},D}$ can be constructed by taking a sequence of $M$ zeros and starting from the $(D+1)$-th most significant bit (i.e., the leftmost one), substituting a $0$ with a $1$, every $D$ bits. If $M=D$, $\mathbf{V}_{\text{lb},D}$ is the all-zero string of length $M$.

Let us clarify the definition considering an example with $M=5$. To each value of $D$ in the set $\{0, 1, 2, 3, 4, 5\}$ corresponds a different vector $\mathbf{V}_{\text{lb},D}$: $\mathbf{V}_{\text{lb},0}=[1 1 1 1 1]$ , $\mathbf{V}_{\text{lb},1}=[0 1 0 1 0]$, $\mathbf{V}_{\text{lb},2}=[0 0 1 0 0]$, $\mathbf{V}_{\text{lb},3}=[0 0 0 1 0]$, $\mathbf{V}_{\text{lb},4}=[0 0 0 0 1]$ and $\mathbf{V}_{\text{lb},5}=[0 0 0 0 0]$.\\

\begin{defn}  We define $\Psi(t)=\sum_{n=1}^tv(n)$ and $\Psi_{\text{lb},D}(t)=\sum_{n=1}^tv_{\text{lb},D}(n)$, where $v(n)$ and $v_{\text{lb},D}(n)$ are the $n$-th bits, starting from the most significant ones, of $\mathbf{V}$ (tentative allocation vector returned by recursion (\ref{eqn:lb_algo})) and $\mathbf{V}_{\text{lb},D}$ (see Definition 1), respectively. In other words, $\Psi(t)$ and $\Psi_{\text{lb},D}(t)$ are the cumulative sum, from left, of the vectors $\mathbf{V}$ and $\mathbf{V}_{\text{lb},D}$, respectively, up to the $t$-th coordinate.\end{defn}

With reference to the example in Fig. \ref{fig:allocation_1}, we have $\Psi(1),\ldots, \Psi(5)= 1, 2, 2, 2, 3$. For $D=2$, we have $\mathbf{V}_{\text{lb},2}=[0 0 1 0 0]$, and $\Psi_{\text{lb},2}(1),\ldots, \Psi_{\text{lb},2}(5)=0,0,1,1,1$.\\

\emph{Theorem 1} Given the allocation vector $\mathbf{V}$ returned by recursion (\ref{eqn:lb_algo}), a  maximum delay less than or equal to $D^*$ is achievable if the following holds: $\Psi(t)\geq \Psi_{\text{lb},D^*}(t)$, $\forall t\in\{1,\ldots,M\}$.\\

\emph{Proof}
 We recall that $\Psi_{\text{lb},D}(t)$ is the total number of $1$'s among the leftmost $t$ bits of the sequence $\mathbf{V}_{\text{lb},D}$ (see Definition 1), while $\Psi(t)$ is the total number of $1$'s among the leftmost $t$ bits of the sequence $\mathbf{V}$. $\Psi(t)\geq \Psi_{\text{lb},D}(t)$, $\forall t\in\{1,\ldots,M\}$, implies that $\mathbf{V}$ has at least as many $1$'s as $\mathbf{V}_{\text{lb},D}$ among the leftmost $t$ positions, $\forall t\in\{1,\ldots,M\}$, which, in turn, implies that $\mathbf{V}$ achieves a maximum delay that is no greater than $D^{*}$, which concludes the proof.\\

In order to find the minimum possible maximum delay starting from a given sequence $\mathbf{V}$, one can start with a delay $D^*=0$ and check if the condition of Theorem $1$ is satisfied. If not, the maximum delay is increased by $1$, and so on.
 %

\begin{algorithm}[htbp]
\caption{\footnotesize \texttt{Min\_Del\_Max\_Rate($\mathbf{V}$)}}
\label{alg:min_del_max_rate}
{\fontsize{10}{10}\selectfont\begin{algorithmic} 
\State $M$ = length($\mathbf{V}$)
\If{$\mathbf{V}==[0, \ldots, 0]$}
\text{ // if no packet can be decoded return the all zero sequence}
\State $\mathbf{S}_{\text{opt}}=[0, \ldots, 0]$
\State \Return $\mathbf{S}_{\text{opt}}$
\EndIf
\State $D$, $k$ = 0

\While{found $== 0$}
\State found $=1$
\State $\mathbf{V}_{\text{lb},D} = [0, \ldots, 0]$ // \text{vector of $M$ zeros}

\For{$i = 1$ to $\left\lfloor \frac{M}{D+1} \right\rfloor$}

\State $\mathbf{V}_{\text{lb},D}[i(D+1)] = 1$
\text{// assign 1 to the $i(D+1)$-th component}
\EndFor
\State cumsum$\_$d $ = 0$
\State cumsum$\_$lb $ = 0$
\For{$j = 1$ to $M$}
\State cumsum$\_$d = cumsum$\_$d$ + \mathbf{V}[j]$
\text{ // calculate $\Psi(j)$}
\State cumsum$\_$lb = cumsum$\_$lb$ + \mathbf{V}_{\text{lb}}[j]$
\text{ // calculate $\Psi_{\text{lb},D}(j)$}
\If{cumsum$\_$d $ < $ cumsum$\_${lb}}
\text{// if cumulative sum is lower, start again increasing delay}
\State found $=0$
\State \textbf{exit} \textbf{for}
\EndIf
\EndFor
\If{found $== 1$}
\State $D^{\text{max}}_{\text{IT}} = D$
\State \textbf{exit} \textbf{while}
\EndIf
\State $D = D + 1$
\EndWhile
\State $\mathbf{S}_{\text{opt}} = \mathbf{V}_{\text{lb},D^{\text{max}}_{\text{IT}}}$
\State excess$\_0 = $ sum($\mathbf{V})-$ sum($\mathbf{V}_{\text{lb},D}$)
\While{$k < $ excess$\_0$} \text{ // assign 1 to the rightmost excess $\_0$ zeros of $\mathbf{V}_{\text{lb},D^{\text{max}}_{\text{IT}}}$}
\If{$\mathbf{S}_{\text{opt}}[M-k] == 0$}
\State $\mathbf{S}_{\text{opt}}[M-k] = 1$
\State $k = k + 1$
\EndIf
\EndWhile

\State \Return $\mathbf{S}_{\text{opt}}$
            \end{algorithmic}}
\end{algorithm}

Using Theorem $1$, the \texttt{Min\_Del\_Max\_Rate} algorithm (Algorithm \ref{alg:min_del_max_rate}) has been obtained. The algorithm takes as input the vector $\mathbf{V}$, which is obtained using the recursion in Eqn. (\ref{eqn:lb_algo}). First the algorithm calculates the minimum achievable maximum delay $D^{\text{max}}_{\text{IT}}$ (see Theorem $1$ and the following note) and derives the vector $\mathbf{V}_{\text{lb},D^{\text{max}}_{\text{IT}}}$. Then it calculates the difference in the number of ones between $\mathbf{V}$ and $\mathbf{V}_{\text{lb},D^{\text{max}}_{\text{IT}}}$ (\verb"excess_0" in the algorithm). By definition of $D^{\text{max}}_{\text{IT}}$, \verb"excess_0" is greater than or equal to zero. Using $\mathbf{V}_{\text{lb},D^{\text{max}}_{\text{IT}}}$ as an initialization allocation vector, the vector $\mathbf{S}_{\text{opt}}$ is then constructed by simply substituting the rightmost  \verb"excess_0" zeros with ones.  The output of the algorithm is the set of messages $\mathbf{S}_{\text{opt}}$ (containing a $1$ or a $0$ in position $t$ if message $W_t$ is to be transmitted, or not) that constitutes the optimal transmission choice in terms of both throughput and maximum delay. It can be easily shown that Algorithm \ref{alg:min_del_max_rate} has a complexity which is quadratic in $M$.

In order to clarify the procedure just described, let us consider again the example in Fig. \ref{fig:allocation_1}. The recursion in Eqn. (\ref{eqn:lb_algo}) returns the vector $\mathbf{V} = [1 1 0 0 1]$, which corresponds to $\Psi = [1 2 2 2 3]$. The algorithm starts with a tentative delay $D^{\text{max}}_{\text{IT}} = 0$, and generates the corresponding sequence $\mathbf{V}_{\text{lb},0} = [1 1 1 1 1]$, with $\Psi_{\text{lb},0} = [1 2 3 4 5]$. Since the condition of Theorem $1$ is not satisfied ($\Psi(3)< \Psi_{\text{lb},0}(3))$, a minimum maximum delay $D^{\text{max}}_{\text{IT}} = 0$ cannot be achieved, and the tentative delay is increased by $1$, i.e., $D^{\text{max}}_{\text{IT}}  = 1$. The corresponding sequences $\mathbf{V}_{\text{lb},1} = [0 1 0 1 0]$ and $\Psi_{\text{lb},1}=[ 0 1 1 2 2]$ are then calculated. The cumulative function $\Psi_{\text{lb},1}$ satisfies the condition of Theorem $1$, which implies that the minimum achievable maximum delay is $D^{\text{max}}_{\text{IT}} = 1$. At this point the algorithm calculates the optimal allocation vector. First, the difference in the number of ones between vector $\mathbf{V}_{\text{lb},1}$ and vector $\mathbf{V}$ (\verb"excess_0") is computed, which in the example is equal to \verb"excess_0"=$1$. Finally, the rightmost \verb"excess_0" zeros in $\mathbf{V}_{\text{lb},1}$ are set to $1$, which leads to the allocation sequence $\mathbf{S}_{\text{opt}}= [0 1 0 1 1]$.
\vspace{-.1in}
\section{Transmission Schemes}\label{sec:enco_schemes}
In this section we introduce four different transmission schemes based on time-sharing. Each channel block is divided among the messages for which the deadline has not yet expired. Thus, while the first channel block is divided among all the messages $W_1,\ldots,W_M$, the second channel block is divided among messages $W_2,\ldots,W_M$, as the deadline of message $W_1$ expires at the end of the first block. In general the encoder divides channel block
$t$ into $M-t+1$ portions $\alpha_{t t},\ldots , \alpha_{M t}$, such that
$\alpha_{mt}\geq0$ and $\sum_{m=t}^{M}\alpha_{mt}=1$. In channel block
$t$, $\alpha_{mt}n$ channel uses are allocated for the transmission
of message $W_m$. We assume that Gaussian codebooks are used in each portion for each message, and the corresponding codelengths are sufficient to achieve the instantaneous capacity.
Then the total amount of received mutual information relative to
message $W_m$ is:
\begin{eqnarray}\label{eqn:tot_mutual_info_time div}
I_{m}^{\text{tot}}\triangleq\sum_{t=1}^m\alpha_{mt}C_t.
\end{eqnarray}
The proposed schemes differ in the way the channel uses are allocated among the messages for which the deadline has not yet expired. Different time allocations lead to different
average throughput and average maximum delay performances.
\vspace{-.1in}
\subsection{Memoryless Transmission (MT)}\label{sec:memoryless}
In \emph{memoryless transmission (MT)} each message is transmitted only within the channel block just before its expiration, that is, message $W_t$ is transmitted over channel block $t$. Equivalently we have $\alpha_{mt}=1$, if $t=m$, and $\alpha_{mt}=0$, otherwise. In MT
message $W_t$ can be decoded if and only if $C_t \geq R$.
Due to the i.i.d. nature of the channel state over blocks, the successful
decoding probability $p\triangleq Pr\{C_t\geq R\}$ is constant over messages.
The probability that exactly $m$ messages are decoded is given by:
\begin{eqnarray}\label{eqn:prob_1_decode_m}
\eta(m) \triangleq \binom{M}{m}p^m(1-p)^{M-m}.
\end{eqnarray}
The average number of decoded messages for the MT scheme is $\overline{T}_{MT}=Mp$.

Next we derive the exact expression for the average maximum delay for MT, denoted by $\overline{D}_{MT}^{\text{max}}$.
The term $Pr\{\text{$D^{\text{max}}\geq d$}\}$ in the summation in Eqn. (\ref{eqn:delay_def_sysmod}) is the probability that a sequence of $M$ Bernoulli random variables with parameter $p$ contains at least $d$ consecutive zeros. This probability can be evaluated by modeling the number of consecutive zeros as a Markov chain, and finding the probability of reaching the final absorbing state of $d$ consecutive zeros. This probability is given in the following theorem:\\

\emph{Theorem 2:}
Let $x_1,\cdots,x_M$ be a sequence of i.i.d. Bernoulli random variables with parameter $p=\mathrm{E}[x_i]$. The probability of having at least $d$ consecutive zeros in the sequence is given by:
\begin{eqnarray}\label{eqn:theorem1}
Pr\{D^{\text{max}}\geq d\} = \sum_{i=0}^k \sum_{r_i=1}^{s_i} a_{d,r_i}\binom{M+r_i-1}{r_i-1}\left(\frac{1}{\varphi_{di}}\right)^M,
\end{eqnarray}
where $k\in \{0,\ldots,M\}$, $k\leq d+1$ is the number of distinct zeros of the polynomial $(1-z)q_d(z)$ where:
\begin{eqnarray}\label{eqn:theorem1_poly00}
q_d(z)=1 - p\sum_{j=1}^{d}z^j(1-p)^{j-1},
\end{eqnarray}
$\varphi_{di}$, $i\in\{0,\ldots,k\}$, are the zeros of $(1-z)q_d(z)$
with multiplicity $s_i$, $a_{d,r_i}$,
$r_i\in\{1,\ldots,s_i\}$, are constants derived from the partial fraction expansion of
\begin{eqnarray}\label{eqn:part_frac}
\frac{(zp)^d}{(1-z)q_d(z)}.
\end{eqnarray}
\emph{Proof:} See Appendix.\\

Finally, by plugging (\ref{eqn:theorem1}) into (\ref{eqn:delay_def_sysmod}) we find:
\begin{eqnarray}\label{eqn:max_del_MT}
\overline{D}^{\text{max}}_{MT}=\sum_{d=1}^M\left[\sum_{i=0}^k \sum_{r_i=1}^{s_i} a_{d,r_i}\binom{M+r_i-1}{r_i-1}\left(\frac{1}{\varphi_{di}}\right)^M\right].
\end{eqnarray}
\vspace{-.3in}
\subsection{Equal Time-Sharing (eTS) Transmission}\label{sec:time_share}
In the equal time-sharing (eTS) transmission scheme each channel block is equally divided among all the messages whose deadline has not expired yet, that is, for
$m=1,\ldots,M$, we have $\alpha_{mt}=\frac{1}{M-t+1}$ for
$t=1,\ldots,m$, and $\alpha_{mt}=0$, for $t=m+1, \ldots, M$.

In eTS, messages whose deadlines are later in time are
allocated more resources; and hence, are more likely to be decoded.
We have $I_{i}^{\text{tot}}<I_{j}^{\text{tot}}$ for $1\leq i<j\leq M$. Hence, the
probability of decoding exactly $m$ messages is:
\begin{eqnarray}\label{eqn:p_dec_m_super}
\eta(m) \triangleq Pr\{I_{m}^{\text{tot}}\geq R \geq I_{m-1}^{\text{tot}}\},
\end{eqnarray}
for $m=0, 1, \ldots, M$, where we define $I_0^{\text{tot}} = 0$ and
$I_{M+1}^{\text{tot}} = \infty$. Since the decoded messages in eTS are always
the last ones,
we can express the average maximum delay of eTS, $\overline{D}_{\text{eTS}}^{\text{max}}$, as a function of its average throughput $\overline{T}_{\text{eTS}}$ as follows:
\begin{align}\label{eqn:rewrite_mutual_info_time div2}
\overline{D}_{\text{eTS}}^{\text{max}} &\triangleq \sum_{m=0}^{M}(M-m) \cdot\eta(m) \notag\\
&= \sum_{m=0}^{M} M \cdot \eta(m) -\sum_{m=0}^{M}m \cdot\eta(m)\notag \\
&= M \left(1- \frac{\overline{T}_{\text{eTS}}}{R} \right).
\end{align}
The numerical analysis of eTS, together with other schemes is presented in Section \ref{sec:num_res}.
\vspace{-.1in}
\subsection{Pre-Buffering (PB) Transmission}
In most practical streaming systems the receiver first accumulates GOPs in the playout buffer and then starts displaying them at a constant frame rate once a sufficient portion of the video has been received, in order to compensate for the delay jitter of arriving packets \cite{luan_2010_impact_video_QoE}. We consider a slightly different version of this type of streaming transmission in which only the last $B$ messages are transmitted while the first packets are not transmitted at all. The first $M-B+1$ channel blocks are used to convey information relative to the last $B$ packets as explained in the following. We call this method \emph{pre-buffering (PB)} transmission.

The initial buffering phase introduces a start-up delay of $M-B$ channel blocks. On the other hand, if a sufficiently large buffering period is chosen, all the transmitted messages can be received correctly, achieving an average throughput of $R\frac{B}{M}$. Transmitted messages are encoded with equal time allocation over the first $M-B+1$ blocks. Due to the delay constraint, message $W_{M-B+1}$ is transmitted up
to channel block $M-B+1$. Hence, in block $M-B+2$ the last
$B-1$ messages are transmitted with equal time allocation. The
process continues up until channel block $M$, in which only message
$W_M$ is transmitted. Next we indicate with $\overline{T}_{\text{PB}}(B)$ and $\overline{D}_{\text{PB}}^{\text{max}}(B)$  the average throughput and the average maximum delay achieved by the scheme using a buffering period of $B$ channel blocks, respectively.
The number $B_{\text{opt}}$ of messages to be transmitted is chosen so that
\begin{eqnarray}\label{eqn:W_opt_sts0}
B_{\text{opt}}=\argmin_{B\in\{1,\ldots,M\}}\left\{\overline{D}^{\text{max}}(B)\right\}.
\end{eqnarray}
Next we show that the $B_{\text{opt}}$, as defined in Eqn. (\ref{eqn:W_opt_sts0}), also maximizes the average throughput. The average throughput when transmitting only the last $B$ messages is given by:
\begin{align}\label{eqn:avg_dec_sTS}
\overline{T}_{\text{PB}}(B) &= \frac{R}{M} \sum_{m=1}^{B} Pr\left\{\text{decode at least $m$ messages} \right\} \notag\\
    &= \frac{R}{M} \sum_{m=1}^{B}Pr\left\{I^{\text{tot}}_{M-m+1}\geq R\right\},
\end{align}
where the mutual information accumulated by the receiver for message $W_m$, for $m = M~-~B~+~1,$ $M~-~B~+~2, \ldots, M$, is given by:
\begin{eqnarray}\label{eqn:tot_mutual_info_time div_windowed}
I_{m}^{\text{tot}}=\frac{1}{B}\sum_{t=1}^{M-B+1}C_t
+ \sum_{t=M-B+2}^{m}\frac{C_t}{M-t+1}.
\end{eqnarray}
From Eqn. (\ref{eqn:avg_dec_sTS}) we have:
\begin{eqnarray}\label{eqn:W_opt_sts2}
\overline{T}_{\text{PB}}(B) &=& \frac{R}{M}\left[B-\sum_{m=1}^{B}Pr\left\{I^{\text{tot}}_{M-m+1}<R\right\}\right] \notag\\
&=& \frac{R}{M}\left[B-\sum_{m=1}^{B}Pr\left\{D^{\text{max}}\geq M-m+1\right\}\right].
\end{eqnarray}
The average maximum delay when only the last $B$ messages are transmitted is:
\begin{eqnarray}\label{eqn:W_opt_sts3}
&\overline{D}_{\text{PB}}^{\text{max}}(B)=M- B +\sum_{d=1}^{B} Pr\left\{D^{\text{max}}\geq M-B+d\right\}.&
\end{eqnarray}
From (\ref{eqn:W_opt_sts2}) and (\ref{eqn:W_opt_sts3}) we find
$$\overline{T}_{\text{PB}}(B)= R\left(1-\frac{\overline{D}^{\text{max}}(B)}{M}\right),$$
and finally
\begin{eqnarray}\label{eqn:W_opt_sts5}
\argmin_{B\in\{1,\cdots,M\}}\left\{\overline{D}_{\text{PB}}^{\text{max}}(B)\right\}=\argmax_{B\in\{1,\cdots,M\}}\left\{\overline{T}_{\text{PB}}(B)\right\}.
\end{eqnarray}
This proves that the average throughput and the maximum delay can be optimized simultaneously. It is not straightforward to come up with an analytical expression for the optimal value of $B$ in the PB scheme for the general case. In the following theorem we derive the optimal fraction of messages $\alpha_{\text{opt}}=\frac{B_{\text{opt}}}{M}$, such that almost all of the transmitted messages can be decoded with probability that approaches $1$ asymptotically as $M$ goes to infinity, if a fraction $\alpha'<\alpha_{\text{opt}}$ of the messages is transmitted, while a fraction smaller than $\alpha_{\text{opt}}$ of the messages can be decoded if $\alpha'>\alpha_{\text{opt}}$.

\emph{Theorem 3} Average throughput of $\alpha R$ can be achieved in the limit of infinite $M$ by transmitting $\alpha M + o(M)$ messages as long as $$\alpha<\alpha_{\text{opt}}\triangleq\frac{1}{\frac{R}{\overline{C}}+1}.$$
If $\alpha>\alpha_{\text{opt}}$, the achieved average throughput is smaller than $\alpha_{\text{opt}}R$.

\emph{Proof} Assume that the last $B$ messages, i.e., $W_{M-B+1},\ldots,W_M$, are transmitted, with $B=M\alpha+o(M)$, $\alpha\leq 1$. Message $W_{M-B+1}$, for which the deadline expires first, is the one that accumulates the least amount of mutual information, that is:
\begin{eqnarray}\label{eqn:theorem52_1}
I_{M-B+1}=\frac{1}{B}\sum_{t=1}^{M-B+1}C_t.
\end{eqnarray}
The probability of decoding all the transmitted messages is then:
\begin{eqnarray}\label{eqn:theorem52_2}
&Pr\left\{I_{M-B+1}\geq R\right\}=Pr\left\{\frac{1}{B}\sum_{t=1}^{M-B+1}C_t\geq R\right\}&\notag\\
&=Pr\left\{\sum_{t=1}^{M-B+1}\frac{C_t}{M-B+1}-\overline{C}\geq \frac{B}{M-B+1}R-\overline{C}\right\}&\notag\\
&=Pr\left\{S_{M-B+1}-\overline{C}\geq \frac{B}{M-B+1}R-\overline{C}\right\},&
\end{eqnarray}
where $S_{M-B+1}\triangleq \sum_{t=1}^{M-B+1}\frac{C_t}{M-B+1}$, is the sample mean of the instantaneous channel capacity over the first $M-B+1$ channel blocks. From the law of large numbers it follows that:
\begin{eqnarray}\label{eqn:theorem52_3}
\lim_{M\rightarrow \infty}Pr\left\{\left|S_{M\left(1-\alpha-\frac{o(M)}{M}\right)}-
\overline{C}\right|>\delta\right\}=0, \ \forall \delta>0.
\end{eqnarray}
Using equations (\ref{eqn:theorem52_2}) and (\ref{eqn:theorem52_3}) we find:
\begin{align}\label{eqn:theorem52_4}
\lim_{M\rightarrow \infty}Pr\left\{I_{M-B+1}\geq R\right\}=
\begin{cases}
1,  \text{ if } \lim_{M\rightarrow \infty} \frac{B}{M-B+1}R<\overline{C} \\
0,  \text{ if } \lim_{M\rightarrow \infty} \frac{B}{M-B+1}R>\overline{C}.
\end{cases}
\end{align}
We can write:
\begin{eqnarray}\label{eqn:theorem52_5}
\lim_{M\rightarrow \infty} \frac{B}{M-B+1}R&=&\lim_{M\rightarrow \infty} \frac{M\alpha+o(M)}{M-M\alpha+o(M)}R\notag\\
&=&\frac{\alpha}{1-\alpha}R.
\end{eqnarray}
Finally, using Eqn. (\ref{eqn:theorem52_5}) in Eqn. (\ref{eqn:theorem52_4}) we find:
\begin{align}\label{eqn:theorem52_6}
\lim_{M\rightarrow \infty}Pr\left\{I_{M-B+1}\geq R\right\}=
\begin{cases}
1,  \text{ if } \alpha<\alpha_{\text{opt}} \\
0,  \text{ if } \alpha>\alpha_{\text{opt}}.
\end{cases}
\end{align}
Eqn. (\ref{eqn:theorem52_6}) implies that if a fraction of messages $\alpha'$ larger than $\alpha_{\text{opt}}$ is transmitted, then the average throughput is less than $\alpha_{\text{opt}}R$, which concludes the proof.\\

\vspace{-.1in}
In Section \ref{sec:num_res}, we provide a numerical optimization of the PB scheme, and compare it with the other proposed transmission strategies and the upper bound. As we will see from the numerical results, this buffering approach can improve the average throughput significantly as it provides rate adaptation at the packet level by eliminating some of the packets, thus increasing the correct decoding probability of the remaining packets.
\vspace{-.1in}
\subsection{Windowed Time Sharing (wTS)}\label{sec:win_tx}
We have seen in the PB scheme that transmitting only a subset of the messages can improve the system throughput by allowing rate adaptation at the packet level. However, in the PB scheme only the last $B$ packets are transmitted leading to a minimum delay of $M-B$ channel blocks. In the next scheme, called the windowed time-sharing (wTS) scheme, $\lceil M/B\rceil$ messages are transmitted, where $\lceil x\rceil$ is the smallest integer greater than or equal to $x$; however,
unlike in PB, the transmitted messages are distributed among the whole set of available messages, that is, only one from $B$
consecutive packets is transmitted over $B$
consecutive channel blocks. So, for instance, if $B=3$, the first
message to be transmitted is $W_3$, which is repeated over channel
blocks $1$, $2$ and $3$, followed by message $W_6$, which is
transmitted in the next three channel blocks, and so on.

The parameter $B$ can be
optimized according to two different criteria, namely to maximize the average throughput or to minimize the delay, which leads to the two variants of the wTS scheme, which we call \emph{throughput-wTS (T-wTS)} and \emph{delay-wTS (D-wTS)}, respectively. In wTS a message is decoded with probability $p_B$ given below:
\begin{align}\label{eqn:wTS}
p_B=Pr\left\{I_{kB}\geq R\right\}=Pr\left\{\sum_{t=kB-W+1}^{\min\{kB,M\}}C_k\geq R\right\},
\end{align}
for $k\in\{1,\ldots,\left\lceil\frac{M}{B}\right\rceil\}$.
A lower bound on
$\overline{D}^{\text{max}}_{wTS}$ can be found by substituting $\left\lfloor\frac{M}{B}\right\rfloor$
for $M$ in Eqn. (\ref{eqn:max_del_MT}), $p_B$ for $p$ in equations (\ref{eqn:theorem1_poly00}) and (\ref{eqn:part_frac}) and multiplying Eqn. (\ref{eqn:max_del_MT}) with $B$. An upper bound can be found in a similar way by using $\left\lceil\frac{M}{B}\right\rceil$ instead of $\left\lfloor\frac{M}{B}\right\rfloor$. Similarly, an upper and a lower bound on $\overline{T}_{wTS}$ are given by $\left\lceil\frac{M}{B}\right\rceil\cdot p_B$ and $\left\lfloor\frac{M}{B}\right\rfloor\cdot p_B$, respectively. Analytical optimization of parameter $B$ in both the T-wTS and D-wTS schemes is elusive and we resort to the numerical analysis presented in the next section.
\vspace{-.1in}
\section{Numerical Results}\label{sec:num_res}
\begin{figure}[]
\centering
\includegraphics[width=3.8in]{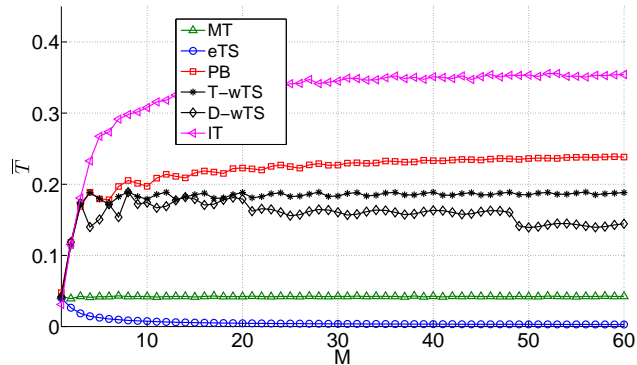}
\caption{\footnotesize Average throughput $\overline{T}$ plotted against the number of
messages transmitted for $SNR=-5$ $\mathrm{dB}$ and $R=1$ bpcu.} \label{fig:through_low}
\end{figure}

\begin{figure}[]
\centering
\includegraphics[width=3.8in]{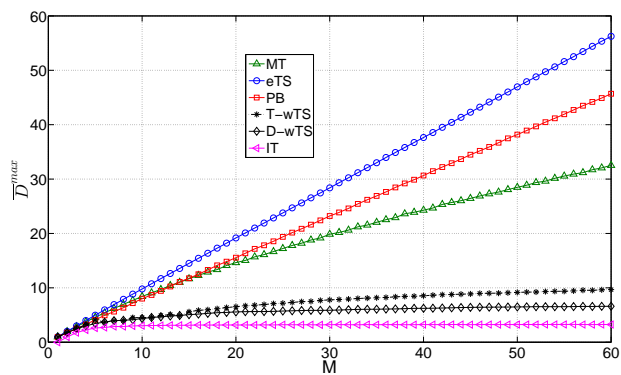}
\caption{\footnotesize Average maximum delay $\overline{D}^{\text{max}}$ plotted against the
number of transmitted messages for $SNR=-5$ $\mathrm{dB}$ and $R=1$ bpcu.}\label{fig:delay_low}
\end{figure}

In this section we compare the average throughput and the average maximum delay of the proposed schemes numerically. The channel model used in the simulations is a Rayleigh block fading channel, in which the channel gain $\phi[t]$ in block number $t$, $t=1,\ldots, M$ (see Eqn. \ref{eqn:capacity}) is a unit-mean exponential random variable that changes in an i.i.d. fashion at the beginning of each channel block and stays constant until the beginning of the next one.
Fig. \ref{fig:through_low} and Fig. \ref{fig:delay_low} show the average throughput and the average maximum delay for the proposed schemes, respectively, for $R=1$ and $\mathrm{SNR}= -5$ $\mathrm{dB}$. Both variants of the wTS scheme perform close to the informed transmitter lower bound in terms of the maximum delay, while the PB scheme is the one with the highest average throughput, followed by T-wTS and D-wTS. The eTS scheme shows quite poor performance in terms of both the delay and the throughput. From the plots it emerges that wTS in its two variants T-wTS and D-wTS, can help to reduce the inter-decoding delay while achieving a relatively good average throughput in the low SNR regime. The transmitter can choose between the two schemes based on its preference between higher throughput and lower inter-decoding delay. While PB provides the highest throughput among the proposed schemes, its inter-decoding delay is significantly high, due to the initial buffering time. PB might be a particularly attractive choice for video streams of long duration, for which the users would be willing to have a larger startup delay to enjoy a higher throughput for the rest of the video.

\begin{figure}[]
\centering
\includegraphics[width=3.8in]{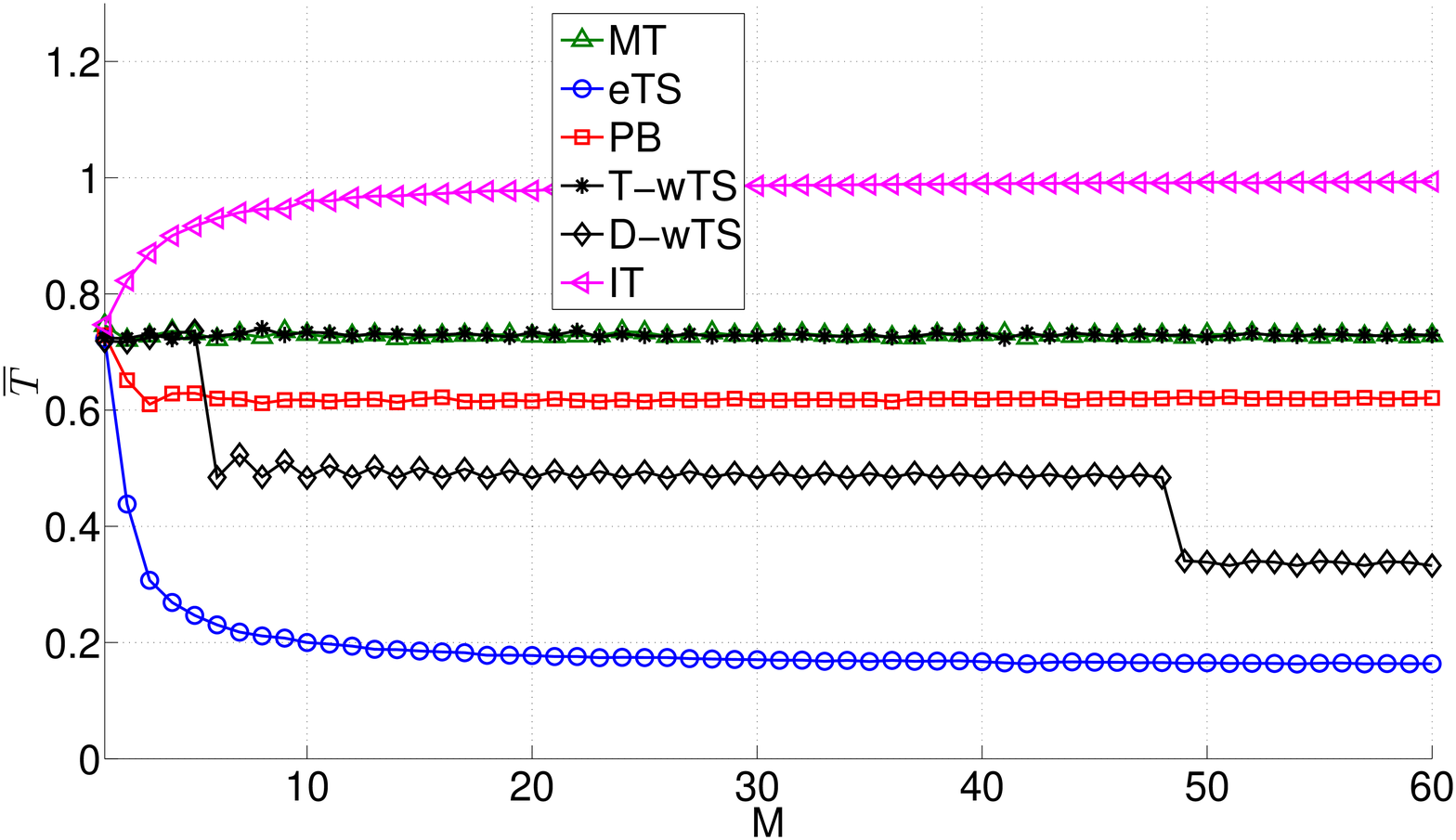}
\caption{\footnotesize Average throughput $\overline{T}$ plotted against the number of
messages transmitted for $SNR=5$ $\mathrm{dB}$ and $R=1$ bpcu.} \label{fig:through_high}
\end{figure}

\begin{figure}[]
\centering
\includegraphics[width=3.8in]{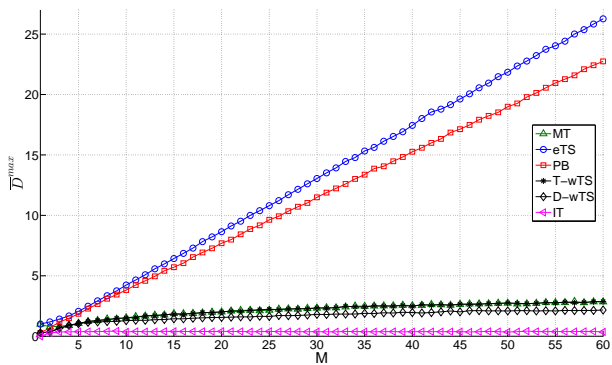}
\caption{\footnotesize Average maximum delay $\overline{D}^{\text{max}}$ plotted against the
number of transmitted messages for $SNR=5$ $\mathrm{dB}$ and $R=1$ bpcu.} \label{fig:delay_high}
\end{figure}

Fig. \ref{fig:through_high} and Fig. \ref{fig:delay_high} show the average throughput and the average maximum delay, respectively, for the proposed schemes for $R=1$ and $\mathrm{SNR}= 5$ $\mathrm{dB}$. Also for this SNR level the two variants of the wTS scheme perform close to the informed transmitter lower bound in terms of maximum delay. The highest average throughput is achieved by the T-wTS scheme together with the MT scheme, followed by the PB, D-wTS and eTS schemes. From Fig. \ref{fig:through_high} and Fig. \ref{fig:delay_high} we see that, when the SNR is high, the MT scheme, together with the T-wTS scheme, achieves the best performances in terms of both delay and average throughput. This suggests that a simple memoryless approach is sufficient when the channel SNR is sufficiently high, while at low SNR more complex encoding techniques can help to significantly improve the performance.
The D-wTS scheme shows a sudden decrease in the average throughput, which, with reference to Fig. \ref{fig:through_high}, also corresponds to a decrease in the slope of the curve at points corresponding to $M=7$ and $M=48$. This is due to the optimization of the window size $B$. We recall that in D-wTS the window size represents the number of channel blocks dedicated to a  message, and is optimized so as to achieve the minimum average maximum delay. While a large $B$ leads to a high decoding probability, it implies a small number of transmitted messages, which bounds from below the minimum delay by $B$. As a matter of fact, only $\lceil \frac{M}{B}\rceil$ messages are transmitted in the wTS scheme, which implies that the maximum delay, in a given realization, is a multiple of $B$. If, for instance, $B=2$ and $m=3$ consecutive messages are lost, the corresponding delay is $m\cdot B=6$. Formally, given a window size $B^*$ there is a certain probability $p_{B^*}^{l}$ of not decoding a message. For any fixed $m\in\{0,\ldots,M\}$, using Eqn. (\ref{eqn:theorem1}) it can be easily shown that the probability of losing at least $m$ consecutive messages increases with $M$.
 Thus a value $B^*$ which is optimal for a certain $M$, may not be the optimal for a larger number of messages, as the probability that more than one consecutive messages get lost increases with $M$. The optimal choice may be to increase $B$, so that the probability of losing consecutive messages is decreased.
 \begin{figure}[h!]
\centering
\includegraphics[width=3.8in]{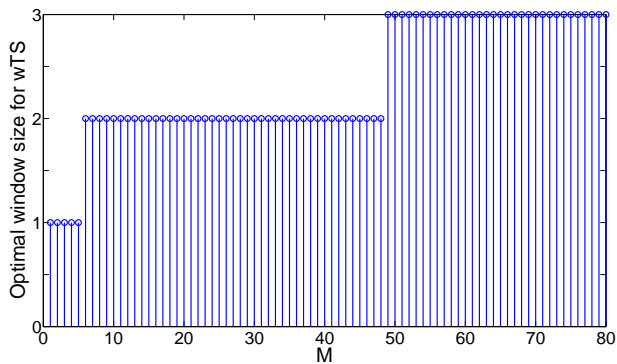}
\caption{Optimal window size ($B$) for the T-wTS scheme plotted versus the total number of messages ($M$) for $SNR=5$ $dB$.} \label{fig:W_opt_wTS}
\end{figure}
This is confirmed by Fig. \ref{fig:W_opt_wTS}, where the optimal window size, obtained numerically, is plotted against the total number of messages.  An increase in $B$ implies a decrease in the slope of the average number of decoded messages, since a smaller fraction of messages is transmitted, as shown in the plots.
 The T-wTS scheme, in which $B$ is optimized so as to achieve the maximum average throughput, shows a good tradeoff between the average throughput, which, unlike D-wTS, is almost independent of the number of messages, and the average maximum delay, performing close to the D-wTS scheme.

 \begin{figure}[]
\centering
\includegraphics[width=3.8in]{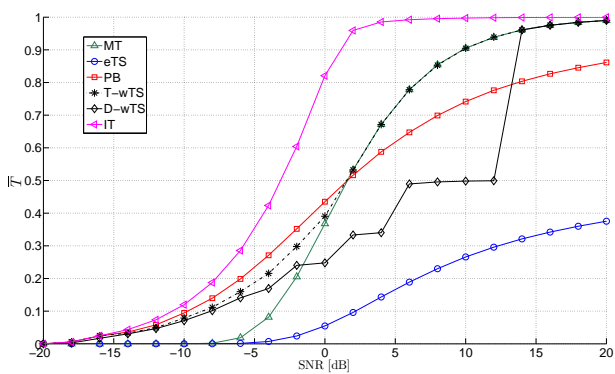}
\caption{\footnotesize Average throughput $\overline{T}$ plotted against the $SNR$ for $M=40$ packets and $R=1$ bpcu.} \label{fig:through_snr}
\end{figure}
\begin{figure}[]
\centering
\includegraphics[width=3.8in]{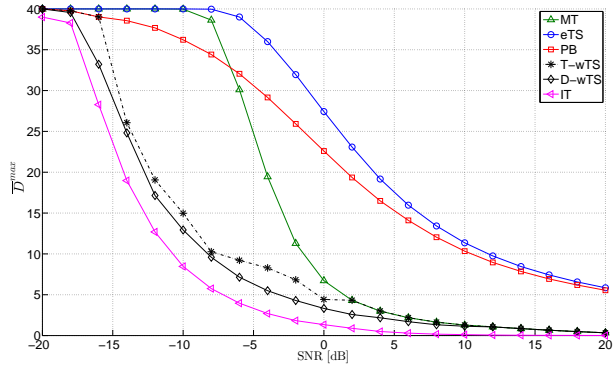}
\caption{\footnotesize Average maximum delay $\overline{D}^{\text{max}}$ plotted against the $SNR$ for $M=40$ packets and $R=1$ bpcu.} \label{fig:delay_snr}
\end{figure}

In Figures \ref{fig:through_snr} and \ref{fig:delay_snr} the average throughput and the average maximum delay, respectively, are plotted against average $SNR$. The plots were obtained for $M=40$ packets and $R=1$ bpcu. As observed in Figures 4 and 6, for $M=40$, the PB scheme outperforms all other schemes in terms of throughput at low SNR (lower than 2 dB), while T-wTS and MT achieve almost the same performance, and outperform the PB scheme at higher $SNR$s. From the figures we observe that the PB scheme is the most robust one against packet losses at low $SNR$, while at higher $SNR$ it is outperformed by all the schemes but the trivial MT. In terms of maximum delay, PB shows relatively poor performance for most of the considered $SNR$ range, which is due to the initial buffering phase. Note that, if, unlike assumed in this paper, the loss of large consecutive chunks of the content were not an issue, and choppy playback were to be avoided, the PB scheme would be the best among the considered schemes since it guarantees that, once the buffering phase is finished, no additional packet is lost, as proven in Theorem 3 for the asymptotic case.
\vspace{-.1in}
\section{Conclusions}\label{sec:conclusions}
We have studied the streaming of stored video content over slow fading channels with per-packet delay constraints. In addition to the classical throughput metric, we have also considered the inter-decoding delay, i.e., the number of consecutive video GOPs that cannot be decoded successfully, as a performance measure. We have proposed four different transmission schemes based on time-sharing. We have carried out theoretical as well as numerical analysis for the average throughput and maximum delay performances. We have also derived bounds on both the average throughput and maximum inter-decoding delay by introducing an informed transmitter bound, in which the transmitter is assumed to know the channel states in advance. We have seen that the wTS scheme can provide a good trade-off between the average throughput and the maximum inter-decoding delay by deciding on the proportion of transmitted video packets. In practice this corresponds to reducing the coding rate of the video at the packet level. We have also proved that in the PB scheme almost all transmitted messages can be decoded with a probability that goes to $1$ as $M$ goes to infinity if only a fraction of the messages smaller than a threshold value, which depends on the transmission rate and the average channel capacity, are transmitted.
\vspace{-.1in}
\section*{Appendix}
\subsection*{Proof of Theorem 1}
\begin{figure}[h!]
\centering
\includegraphics[width=3.4in]{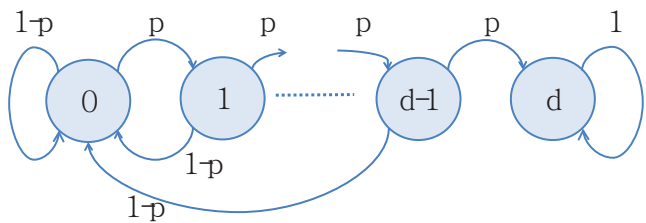}
\caption{\footnotesize Markov chain for the calculation of the average maximum delay in memoryless transmission.} \label{fig:z_transform}
\end{figure}

The probability of having a run of at least $d$, $d\in \{0,\ldots,M\}$, consecutive zeros in the sequence is equivalent to finding the probability of state $d$ after $M$ steps in the Markov chain depicted in Fig. \ref{fig:z_transform}.
The state $d$ is an absorbing state, i.e., once the process reaches that state, it remains there with probability $1$. Let $\mathbf{p}_t$ be a $d$-length probability mass function, where $\mathbf{p}_t(i)$, $i=0, \ldots, d$, denotes the probability  of being in state $i$ at step $t$.
The vector $\mathbf{p}_t$ of state occupancy at step $t$ for the Markov chain in Fig. \ref{fig:z_transform} can be obtained as:
\begin{align}\label{eqn:state_occup}
\mathbf{p}_t=\mathbf{p}_{t-1}\mathbf{H}=\mathbf{p}_0\mathbf{H}^t,
\end{align}
where $\mathbf{p}_0=[1\ 0\ \cdots\ 0]$ and $\mathbf{H}$ is the $(d+1)\times (d+1)$ transition matrix of the chain which has the following structure:
\begin{align}\label{eqn:trans_matrix}
\mathbf{H}= \left( \begin{array}{ccccccc}
1-p & p & 0 & 0&\cdots & 0 &0\\
1-p & 0 & p & 0&\cdots & 0&0 \\
\vdots&\vdots&&\vdots&&\vdots\\
1-p & 0 & 0 & 0&\cdots &0& p\\
0 & 0 & 0 & 0&\cdots &0& 1 \end{array} \right).
\end{align}
The probability of being in state $d$ after $M$ steps, $\mathbf{p}_M(d)$, can be found from Eqn. (\ref{eqn:state_occup}). Since $\mathbf{p}_0=[1\ 0\ \cdots\ 0]$ we have:
  \begin{align}\label{eqn:state_occup_element}
\mathbf{p}_M(d)=\mathbf{H}^M(1,d+1).
\end{align}
  In order to evaluate $\mathbf{H}^M(1,d+1)$, we apply the \emph{Z-transform} to Eqn. (\ref{eqn:state_occup}), taking into account that the recursive formula is defined only for $t\geq 1$.
The Z-transform $\mathcal{P}(z)$ of a discrete vector function $\mathbf{p}_t$ is defined as \cite{klein1}:
\begin{align}\label{eqn:z_trans_def}
\mathcal{P}_z\triangleq \mathcal{Z}(\mathbf{p}_t)=\sum_{t=0}^{+\infty}\mathbf{p}_tz^t.
\end{align}
To account for the fact that $t\geq1$ in Eqn. (\ref{eqn:state_occup}) we can write:
\begin{align}\label{eqn:z_trans_apply1}
\sum_{t=1}^{+\infty}\mathbf{p}_tz^t=
\sum_{t=0}^{+\infty}\mathbf{p}_tz^t-\mathbf{p}_0=
\mathcal{P}_z-\mathbf{p}_0,
\end{align}
and
\begin{eqnarray}\label{eqn:z_trans_apply2}
\sum_{t=1}^{+\infty}\mathbf{p}_{t-1}\mathbf{H}z^t&=&z\sum_{t=1}^{+\infty}\mathbf{p}_{t-1}\mathbf{H}z^{t-1}\notag\\
&=&z\sum_{t=0}^{+\infty}\mathbf{p}_{t}\mathbf{H}z^{t}\notag\\&=&z\mathcal{P}_z\mathbf{H}.
\end{eqnarray}
Plugging Eqn. (\ref{eqn:z_trans_apply1}) and Eqn. (\ref{eqn:z_trans_apply2}) into Eqn. (\ref{eqn:state_occup}) we find:
\begin{align}\label{eqn:z_trans_apply3}
\mathcal{P}_z=\mathbf{p}_0\left(\mathbf{I}-z\mathbf{H}\right)^{-1},
\end{align}
where $\mathbf{I}$ is the $(d+1)\times(d+1)$ identity matrix.

The Z-transform $\mathcal{C}_z$ of a matrix $\mathbf{C}_t$ of functions in the discrete variable $t$ is defined as:
\begin{align}\label{eqn:z_trans_def_matr}
\mathcal{C}_z\triangleq\mathcal{Z}(\mathbf{C}_t)=\sum_{t=0}^{+\infty}\mathbf{C}_tz^t.
\end{align}
Note that in  Eqn. (\ref{eqn:z_trans_def_matr}) the term $z^t$ is a scalar function of $z$ and $t$ which is multiplied to each of the elements of matrix $\mathbf{C}_t$.
By comparing Eqn. (\ref{eqn:z_trans_apply3}) with Eqn. (\ref{eqn:state_occup}), we see that $\left(\mathbf{I}-z\mathbf{H}\right)^{-1}$ is the Z-transform of the matrix $\mathbf{H}^t$ having functions in the discrete variable $t$ as elements. We have:
\begin{align}\label{eqn:trans_matrix_transform}
\mathbf{I}-z\mathbf{H}= \left( \begin{array}{ccccccc}
1-z(1-p) & -zp & 0 & 0&\cdots & 0 &0\\
-z(1-p) & 1 & -zp & 0&\cdots & 0&0 \\
\vdots&\vdots&&\vdots&&\vdots\\
-z(1-p) & 0 & 0 & 0&\cdots &1& -zp\\
0 & 0 & 0 & 0&\cdots &0& 1-z \end{array} \right).
\end{align}
Once $\left(\mathbf{I}-z\mathbf{H}\right)^{-1}$ is known, it is sufficient to inversely transform it and get $\mathbf{H}^t$. We find the inverse of matrix (\ref{eqn:trans_matrix_transform}) for a generic $d$ using Gauss-Jordan elimination. As we only need the element $\mathbf{H}^M(1,d+1)$, we only report the first row of $\left(\mathbf{I}-z\mathbf{H}\right)^{-1}$ in Eqn. (\ref{eqn:matrix_row}) at the top of the next page,
\begin{figure*}[t!]
\begin{small}
\begin{align}\label{eqn:matrix_row}
\left(\mathbf{I}-z\mathbf{H}\right)^{-1}_{[1,:]}= \frac{1}{(1-z)q_d(z)}\left[\begin{array}{ccccccc}(1-z)&(1-z)(zp)&(1-z)(zp)^2&\cdots&(1-z)(zp)^{d-1}&(zp)^d\end{array}\  \right].
\end{align}
\end{small}
\hrulefill
\end{figure*}
where
\begin{eqnarray}\label{eqn:inverse_den_poly}
q_d(z)\triangleq 1-p\sum_{j=1}^{d}z^j(1-p)^{j-1}.
\end{eqnarray}
The probability of being in state $d$ at step $M$ is the inverse Z-transform of element $(1,d+1)$ of matrix $\left(\mathbf{I}-z\mathbf{H}\right)^{-1}$, i.e.:
\begin{eqnarray}\label{eqn:probab_in_D}
\mathbf{p}_M(d+1)=\mathcal{Z}^{-1}\left\{\frac{(zp)^d}{(1-z)q_d(z)}\right\}_{t=M},
\end{eqnarray}
where $\mathcal{Z}^{-1}\{\mathcal{P}_z\}$ is the inverse Z-transform of $\mathcal{P}_z$ defined as \cite{klein1}:
\begin{eqnarray}\label{eqn:inverse_z_transf}
\mathcal{Z}^{-1}\{\mathcal{P}_z\}=\frac{-1}{2\pi j}\oint_{\gamma}\mathcal{P}_z z^{-t-1}dz=\mathbf{p}_t,
\end{eqnarray}
$\gamma$ being a counterclockwise-oriented circle around the origin of the complex plane.
An easier way to solve Eqn. (\ref{eqn:probab_in_D}) is to decompose the Z-transform using partial fraction decomposition, i.e., rewriting $\mathcal{P}_z$ as:
\begin{eqnarray}\label{eqn:part_frac_decompose}
\mathcal{P}_z=\frac{(zp)^d}{(1-z)q_d(z)}=\sum_{i=0}^k\sum_{r_i=1}^{s_i} a_{d,r_i}\left(\frac{1}{1-\frac{z}{\varphi_{d,i}}}\right)^{r_i},
\end{eqnarray}
where $\varphi_{d,i}$, $i\in\{0,\ldots,k\}$, are the $k\leq d+1$ distinct zeros with degree $d+1$ and multiplicity $s_i$ of the polynomial $(1-z)q_d(z)$, while $a_{d,r_i}$, $r_i\in\{1,\ldots,s_i\}$, are constants deriving from the partial fraction expansion of $\mathcal{P}_z$.
Once in the form of Eqn. (\ref{eqn:part_frac_decompose}), $\mathcal{P}_z$ can be inversely transformed using the linearity of the inverse Z-transform and the fact that:
\begin{eqnarray}\label{eqn:binom_transf_2}
\mathcal{Z}^{-1}\left\{\left(\frac{1}{1-\frac{z}{\varphi_{d,i}}}\right)^{r_i}\right\}=\binom{t+r_i-1}{r_i-1}\left(\frac{1}{\varphi_{d,i}}\right)^t.
\end{eqnarray}
Eqn. (\ref{eqn:binom_transf_2}) follows from the fact that:
\begin{eqnarray}\label{eqn:binom_transf}
\mathcal{Z}\left\{\left(\frac{1}{\varphi}\right)^t\right\}&\triangleq&\sum_{t=0}^{\infty}\left(\frac{1}{\varphi}\right)^tz^t\notag\\ &=&\sum_{t=0}^{\infty}\left(\frac{z}{\varphi}\right)^t\notag\\&=&\frac{1}{1-z/\varphi},\end{eqnarray}
for $|z|<\varphi$, and from the fact that the Z-transform of the convolution of sequences is the product of the Z-transform of the individual sequences (see \cite[Appendix 1]{klein1} for further details).
Finally, using Eqn. (\ref{eqn:binom_transf}) and Eqn. (\ref{eqn:part_frac_decompose}) and putting $t=M$, we find:
\begin{align}\label{eqn:theorem1_appendix}
Pr\{D^{\text{max}}\geq d\}= & \mathbf{p}_M(d+1)\notag\\=&\sum_{i=0}^k \sum_{r_i=1}^{s_i} a_{d,r_i}\binom{M+r_i-1}{r_i-1}\left(\frac{1}{\varphi_{di}}\right)^M.
\end{align}
\bibliographystyle{ieeebib}
\bibliography{articoli}

\begin{thebibliography}{10}

\bibitem{cisco_forecast_2014}
Cisco~Sysems Inc,
\newblock ``Cisco visual networking index,''
\newblock White paper, Cisco Sysems, Inc, http://www.cisco.com/, Feb. 2014.

\bibitem{ozarow_TVT_mobile_radio}
L.~H. Ozarow, S.~Shamai, and A.~D. Wyner,
\newblock ``Information theoretic considerations for cellular mobile radio,''
\newblock {\em IEEE Trans. on Vehicular Technology}, vol. 43, no. 2, pp.
  359--378, May 1994.

\bibitem{goldsmith_TIT1997_capacity_fading}
A.~J. Goldsmith and P.~P. Varaiya,
\newblock ``Capacity of fading channels with channel side information,''
\newblock {\em IEEE Trans. on Info. Theory}, vol. 43, no. 6, pp. 1986--1992,
  Nov. 1997.

\bibitem{caire_TIT91_power_control}
G.~Caire, G.~Taricco, and E.~Biglieri,
\newblock ``Optimum power control over fading channels,''
\newblock {\em IEEE Trans. on Info. Theory}, vol. 45, no. 5, pp. 1468--1489,
  July 1999.

\bibitem{berry_gallagher_TIT_2002}
R.~A. Berry and R.~G. Gallager,
\newblock ``Communication over fading channels with delay constraints,''
\newblock {\em IEEE Trans. on Info. Theory}, vol. 48, no. 5, pp. 1135--1149,
  May 2002.

\bibitem{zhang_jsac_2010_video_fading}
H.~Zhang, Y.~Zheng, M.~A. Khojastepour, and S.~Rangarajan,
\newblock ``Cross-layer optimization for streaming scalable video over fading
  wireless networks,''
\newblock {\em IEEE J. Sel. Areas Commun.}, vol. 28, no. 3, pp. 344--353, Apr.
  2010.

\bibitem{Hanly:IT:98}
S.~V. Hanly and D.~N.~C. Tse,
\newblock ``Multiaccess fading channels. {II}. {Delay-limited} capacities,''
\newblock {\em IEEE Trans. on Info. Theory}, vol. 44, no. 7, pp. 2816--2831,
  Nov. 1998.

\bibitem{Tse:book}
D.~Tse and P.~Viswanath,
\newblock {\em Fundamentals of Wireless Communication},
\newblock Cambridge University Press, 2005.

\bibitem{gong_twc_2014_delay_video}
C.~Gong and X.~Wang,
\newblock ``Adaptive transmission for delay-constrained wireless video,''
\newblock {\em IEEE Trans. on Wireless Commun.}, vol. 13, no. 1, pp. 49--61,
  Jan. 2014.

\bibitem{ng_gunduz_2007_recursive_power}
C.~T~.K. Ng, D.~G\"{u}nd\"{u}z, A.~J. Goldsmith, and E.~Erkip,
\newblock ``Distortion minimization in {Gaussian} layered broadcast coding with
  successive refinement,''
\newblock {\em IEEE Trans. on Info. Theory}, vol. 55, no. 11, pp. 5074--5086,
  Nov. 2009.

\bibitem{gunduz_2008_joint_source}
D.~G\"{u}nd\"{u}z and E.~Erkip,
\newblock ``Joint source-channel codes for {MIMO} block-fading channels,''
\newblock {\em IEEE Trans. on Info. Theory}, vol. 54, no. 1, pp. 116--134, Jan.
  2008.

\bibitem{Shamai:ISIT:97}
S.~Shamai,
\newblock ``A broadcast strategy for the {Gaussian} slowly fading channel,''
\newblock in {\em IEEE Int. Symp. Info. Theory}, Ulm, Germany, June-July 1997.

\bibitem{drapernework2006benefitsstreaming_fading}
S.~C. Draper and M.D. Trott,
\newblock ``Costs and benefits of fading for streaming media over wireless,''
\newblock {\em IEEE Network}, vol. 20, no. 2, pp. 28--33, Mar. 2006.

\bibitem{wang_video_proc_and_comm}
Y.~Wang, J.~Ostermann, and Y.-Q. Zhang,
\newblock {\em Video processing and communications},
\newblock Prentice Hall, 2002.

\bibitem{sali2007_feedback_implosion}
A.~Sali, G.~Acar, B.~Evans, and G.~Giambene,
\newblock ``Feedback implosion suppression algorithm for reliable multicast
  data transmission over geostationary satellite networks,''
\newblock in {\em Int. Workshop on Satellite and Space Commun.}, Salzburg, Sep.
  2007.

\bibitem{video_quality_assessment_JSTSP_2012}
A.~K. Moorthy, L.~K. Choi, A.~C. Bovik, and G.~de~Veciana,
\newblock ``Video quality assessment on mobile devices: Subjective, behavioral
  and objective studies,''
\newblock {\em IEEE J. of Selected Topics in Signal Processing}, vol. 6, no. 6,
  pp. 652--671, Oct. 2012.

\bibitem{lin_tip2010_model_video_loss}
T.-L. Lin, S.~Kanumuri, Y.~Zhi, D.~Poole, P.~C. Cosman, and A.R. Reibman,
\newblock ``A versatile model for packet loss visibility and its application to
  packet prioritization,''
\newblock {\em IEEE Trans. on Image Processing}, vol. 19, no. 3, pp. 722--735,
  Mar. 2010.

\bibitem{liang_2008_effect_packet_loss}
Y.~J. Liang, J.~G. Apostolopoulos, and B.~Girod,
\newblock ``Analysis of packet loss for compressed video: Effect of burst
  losses and correlation between error frames,''
\newblock {\em IEEE Trans. on Circuits and Syst. for Video Technology}, vol.
  18, no. 7, pp. 861--874, July 2008.

\bibitem{huszak_isccsp2010_gop_loss_effect}
A.~Huszak and S.~Imre,
\newblock ``Analysing {GOP} structure and packet loss effects on error
  propagation in {MPEG-4} video streams,''
\newblock in {\em IEEE Int. Symp. on Commun., Control and Signal Processing
  (ISCCSP)}, Limassol, Cyprus, Mar. 2010.

\bibitem{ITU_T_Y1541}
{International Telecommunication Union-Telecommunication Standardization Sector
  (ITU-T)},
\newblock ``{ITU-T} recommendation {Y.l541}, network performance objectives for
  {IP}-based services,''
\newblock http://www.itu.int/, Dec. 2011.

\bibitem{toni_JSAC2012_ch_cod_optim_video_str}
L.~Toni, P.~C. Cosman, and L.~B. Milstein,
\newblock ``Channel coding optimization based on slice visibility for
  transmission of compressed video over {OFDM} channels,''
\newblock {\em IEEE J. Sel. Areas Commun.}, vol. 30, no. 7, pp. 1172--1183,
  Aug. 2012.

\bibitem{zhao_glob_2010_video_str_ofdm}
S.~Zhao, Y.~Zhang, and L.~Gui,
\newblock ``Optimal resource allocation for video delivery over {MIMO OFDM}
  wireless systems,''
\newblock in {\em IEEE Global Telecommun. Conference {(GLOBECOM)}}, Miami, FL,
  U.S.A., Dec. 2010.

\bibitem{zeng_2012:joint_coding}
W.~Zeng, C.~T.~K. Ng, and M.~Medard,
\newblock ``Joint coding and scheduling optimization in wireless systems with
  varying delay sensitivities,''
\newblock in {\em IEEE Commun. Society Conf. on Sensor, Mesh and Ad Hoc Commun.
  and Networks (SECON)}, Seoul, Korea, June 2012.

\bibitem{Aparicio-Pardo:2015:TLA:2713168.2713177}
R.~Aparicio-Pardo, K.~Pires, A.~Blanc, and G.~Simon,
\newblock ``Transcoding live adaptive video streams at a massive scale in the
  cloud,''
\newblock in {\em ACM Multimedia Systems Conference}, Portland, Oregon, U.S.A.,
  Mar. 2015.

\bibitem{van_der_schaar_2003}
M.~van~der Schaar, S.~Krishnamachari, C.~Sunghyun, and X.~Xiaofeng,
\newblock ``Adaptive cross-layer protection strategies for robust scalable
  video transmission over 802.11 {WLANs},''
\newblock {\em IEEE J. Sel. Areas Commun.}, vol. 21, no. 10, pp. 1752--1763,
  Dec. 2003.

\bibitem{Bogino:ISCAS:07}
M.~C.~O. Bogino, P.~Cataldi, M.~Grangetto, E.~Magli, and G.~Olmo,
\newblock ``Sliding-window digital fountain codes for streaming multimedia
  contents,''
\newblock in {\em Int. Symp. Circuits Systems}, New Orleans, LA, U.S.A., May
  2007.

\bibitem{Badr:Globecom:10}
A.~Badr, A.~Khisti, and E.~Martinian,
\newblock ``Diversity embedded streaming erasure codes ({DE-SCo}):
  Constructions and optimalty,''
\newblock in {\em IEEE Global Telecommun. Conf.}, Miami, FL, U.S.A., Dec. 2010.

\bibitem{Leong:ISIT:12}
D.~Leong and T.~Ho,
\newblock ``Erasure coding for real-time streaming,''
\newblock in {\em IEEE Int. Symp. Info. Theory (ISIT)}, Boston, MA, U.S.A.,
  July 2012.

\bibitem{khisti_11_IT}
A.~Khisti and S.C. Draper,
\newblock ``Streaming data over fading wireless channels: {The
  }diversity-multiplexing tradeoff,''
\newblock in {\em IEEE Int. Symp. Info. Theory}, St. Petersburg, Russia, Aug.
  2011.

\bibitem{cocco11_real_time_BC}
G.~Cocco, D.~G\"{u}nd\"{u}z, and C.~Ibars,
\newblock ``Real-time broadcasting over block-fading channels,''
\newblock in {\em IEEE Int. Symp. Wireless Commun. Syst.}, Aachen, Germany,
  Nov. 2011.

\bibitem{cocco13_twc_streaming}
G.~Cocco, D.~G\"{u}nd\"{u}z, and C.~Ibars,
\newblock ``Streaming transmission over block fading channels with delay
  constraint,''
\newblock {\em IEEE Trans. on Wireless Commun.}, vol. 12, no. 9, pp.
  4315--4327, Aug. 13.

\bibitem{cocco_icc_2013}
G.~Cocco, D.~G\"{u}nd\"{u}z, and C.~Ibars,
\newblock ``Throughput and delay analysis in video streaming over block-fading
  channels,''
\newblock in {\em IEEE Int. Conf. on Commun. (ICC)}, Budapest, Hungary, June
  2013.

\bibitem{toni_TIP2009}
L.~Toni, Y.S. Chan, P.C. Cosman, and L.B. Milstein,
\newblock ``Channel coding for progressive images in a {2-D} time-frequency
  {OFDM} block with channel estimation errors,''
\newblock {\em IEEE Trans. on Image Processing}, vol. 18, no. 11, pp.
  2476--2490, Nov 2009.

\bibitem{thomas_infotheo}
T.~M. Cover and J.~A. Thomas,
\newblock {\em Elements of Information Theory},
\newblock John Wiley \& Sons, second edition edition, 2006.

\bibitem{luan_2010_impact_video_QoE}
T.~H. Luan, L.~X. Cai, and S.~Xuemin,
\newblock ``Impact of network dynamics on user's video quality: Analytical
  framework and {QoS} provision,''
\newblock {\em IEEE Trans. on Multimedia}, vol. 12, no. 1, pp. 64--78, Jan.
  2010.

\bibitem{klein1}
L.~Kleinrock,
\newblock {\em Queueing Systems}, vol.~I,
\newblock John Wiley and Sons, 1975.

\end{thebibliography}
\flushend

\end{document}